\documentclass[article,pra,twocolumn,nofootinbib]{revtex4-1}

\usepackage{amsmath,xcolor}
\usepackage{amsfonts}
\usepackage{amssymb,ulem}
\usepackage{graphicx}
\usepackage{srcltx}

\newcommand{\beq}{\begin{equation}}
\newcommand{\eeq}{\end{equation}}
\newcommand{\ket} [1] {|#1\rangle}
\newcommand{\bra} [1] {\langle#1|}

\newcommand{\M}{\hat{\mathcal M}}
\newcommand{\N}{\hat{\mathcal N}}

\renewcommand{\emph}[1]{{\it#1}}

\begin{document}

\title{Quantum Randomness Certified by the Uncertainty Principle}
%
\author{Giuseppe Vallone}
\author{Davide G. Marangon}
\author{Marco Tomasin}
\author{Paolo Villoresi}
\affiliation{Department of Information Engineering, University of Padova, I-35131 Padova, Italy}



\begin{abstract}
We present an efficient method to extract the amount of true randomness that can
be obtained by a Quantum Random Number Generator (QRNG).
By repeating the measurements of a quantum system and by swapping between two mutually unbiased bases,
a lower bound of the achievable true randomness can be evaluated. The bound is obtained thanks to
the uncertainty principle of complementary measurements applied to min- and max- entropies.
We tested our method with two different QRNGs,
using a train of qubits or ququart, demonstrating the scalability toward practical applications. 
\end{abstract}
%

\maketitle

\section{Introduction}
Random numbers are of fundamental importance for scientific and practical applications. 
On the last years, great effort has been devoted to Quantum Random Number Generators (QRNG), 
based on the intrinsic randomness of the quantum measurement process \cite{jenn00rsi,fior07pra,wei09ope,svoz09pra,furs10ope,jofr11ope,gall13nco,abel14ope,piro10nat,dhar13pra}.
  Theoretical analyses about the security and the real content of randomness have been given only recently \cite{colb06thesis,piro10nat,colb11jpa,vazi12pta,dhar13pra}. 
It has been shown that {\it true random numbers}, 
namely uniform and uncorrelated from any classical or quantum side-information held by an eavesdropper,  
can be achieved by using the randomness \emph{expansion} \cite{vazi12pta,piro13pra} or \emph{amplification} protocols \cite{colb12nph,gall13nco}. 
Expansion refers to a protocol able to generate true random numbers by starting with a short random seed.
In an amplification protocol,
the initial seed can have arbitrarily weak (but nonzero) randomness at the price of lower output rate.
However, both protocols
are very demanding under the experimental point of view since, by operating in the device independent framework, the only way to get perfect randomness is to enforce conditions of no-locality and no-signalling between two parties that violate a (loophole-free) Bell inequality \cite{piro10nat}. 

A general QRNG works as follows: given a  $d$-level quantum system $A$
prepared in a state $\rho_A$, the random variable $Z$ is obtained by measuring the state $\rho_A$
with a $d$ outcome measurement $\mathbb Z$: each outcome $z$ is obtained with a given probability $P_z$.
If the state $\rho_A$ is pure, the number of true random bits that can be extracted from each measurement
is quantified by the classical min-entropy $H_{\infty}(Z)=-\max_z(\log_2P_z)$.
In this work we aim to deal with a generic scenario, in which the state $\rho_A$ is not pure and therefore
the system $A$ is correlated with another quantum system, denoted by $E$. In this case it is necessary to estimate the amount
of (quantum) information that an adversary Eve holding the system $E$ has on the variable $Z$.
The importance of this estimation can be illustrated by a simple example. Let's suppose that Eve holds two entangled photons in the state
$\ket{\Phi}=\frac{1}{\sqrt{2}}(\ket{HH}+\ket{VV})$ and sends to Alice one of the two photons as the system she uses
for the randomness extraction. If Alice measures in the $\{\ket{H},\ket{V}\}$ basis
she obtains a perfect random bit from the point of view of the classical min-entropy, since the two outcomes, $\ket{H}$ and 
$\ket{V}$, are equally probable. However, due to the correlations in the $\ket{\Phi}$ state, Eve knows perfectly the outputs of 
Alice's measurements: the ``random'' bit held by Alice can be predicted with certainty by Eve.

{The amount of true random bits that can be extracted from the random variable $Z$,
if one requires uniformity and independence from {the environment  system $E$},
is given by the conditional min-entropy $H_{\rm min}(Z|E)$ \cite{frau13qph,de12joc}. Indeed, the probability of guessing $Z$
 by holding the quantum system $E$ is given by \cite{koni09ieee}
  \beq
\label{p_guess} 
p_{guess}(Z|E)=2^{-H_\text{min}(Z|E)}\,.
 \eeq
}
For instance, in the previous example with the entangled state $\ket{\Phi}$, $p_{guess}(Z|E)=1$ and
the system held by Alice doesn't allow the generation of true random numbers.
 
We will present a method, based on the Uncertainty Principle (UP),
to estimate the conditional min-entropy and then the amount of true randomness that can be obtained by a given source.
 We will show and experimentally test that, by measuring the system in conjugate observables $\mathbb Z$ and $\mathbb X$,
 it is possible to obtain the following bound on the conditional min-entropy
 \beq\label{simple_bound}
H_{\text{min}}(Z|E)\geq \log_2d-H_{1/2}(X)\,,
\eeq
where $d$ is the dimension of the Hilbert space and $H_{1/2}(X)$ the max-entropy of $\mathbb X$ outcomes (see below).
The measurement $\mathbb Z$ is used to generate the random sequence $Z$,
while the measurement $\mathbb X$ is used to quantify the amount of true-randomness contained in $Z$.
In our protocol we do not use any assumption on the source $\rho_A$:
an adversary, called Eve can have full control on the source and the environment $E$.
{The bound \eqref{simple_bound} is achieved by only assuming trusted measurements device,
meaning that
Eve has no access to it and that 
the device performs a given POVM that are only sensitive to a subspace of dimension
$d$. To prevent  the possibility that an
adversary controls 
the detection efficiency, as reported in quantum hacking against detectors \cite{zhao08pra,lyde10nap,xu10njp}, 
it is necessary to monitor all detector parameters, such as bias voltage, current, and temperature \cite{lyde11pra}.
}
The advantage of the presented method resides on its simplicity: no Bell inequality violation is required but it is only necessary 
to measure the system in two conjugate bases. With an initial seed of true randomness, our protocol 
is able to expand the randomness by taking into account all possible side quantum information possessed by Eve.

\section{Proof of main result}
In this section we derive our main result \eqref{simple_bound}. We first start by reviewing 
the uncertainty relation for min- and max- conditional 
entropies introduced in \cite{toma11prl, rene09prl,bert10nap}.

\subsection{Uncertainty principle}
Let's consider three quantum systems $A$, $B$ and $E$ and $\rho_{ABE}$ a tripartite state.
 Define $\mathbb Z$ and $\mathbb X$ as two POVMs on $A$ with elements $\{\M_z\}$ and $\{\N_x\}$, 
and random outcomes $Z$ and $X$ encoded in two orthonormal bases $\{\ket{z}\}$ and $\{\ket{x}\}$.
Then, the uncertainty principle is written as
 \beq
 \label{UP}
H_{\text{min}}(Z|E)_\rho+H_{\text{max}}(X|B)_\rho\geq q\,,
\eeq
where the min-entropy and max-entropy
(see Appendix \ref{min-entropy} and \cite{koni09ieee} for min- and max- entropy definition)
 are evaluated on the post-measurement states 
$\rho_{ZE}\equiv\sum_z\ket{z}\bra{z}\otimes \text{Tr}_{AB}[\M_z\rho_{ABE}]$,
$\rho_{XB}\equiv\sum_x\ket{x}\bra{x}\otimes \text{Tr}_{AE}[\N_x\rho_{ABE}]$
 and 
 \beq
 q \equiv\log_2\frac1c\,,
\qquad
 c\equiv\text{max}_{z,x} \|\sqrt{\M_z}\sqrt{\N_x}\|^2_\infty\,.
 \eeq
 The parameter $c$ represents the maximum ``overlap'' between the two POVMs and 
 $q$ quantifies the 
 ``incompatibility'' of the measurements.
 If $\M_z$ and $\N_x$ are projective measurements corresponding to 
 Mutually-Unbiased bases in dimension $d$, then $c=\frac{1}{d}$.

\subsection{Proof of the bound}
In a QRNG, 
Alice measures its system $\rho_A$ by using a POVM measurement $\mathbb{Z}\equiv\{\M_z\}$\footnote{We employed 
POVMs to present our method in a general framework, but projective
  measurements are more suited for practical applications.}.
The state $\rho_A$ is in general correlated with an external system $E$
such that $\rho_A={\rm Tr}_E[\rho_{AE}]$.
The possible  outcomes of the POVM can be encoded in an orthonormal basis $\{\ket{z}_A\}$, such
that the post-measurement state is 
$\rho_{ZE}\equiv\sum_z\ket{z}\bra{z}\otimes \text{Tr}_{A}[\M_z\rho_{AE}]=
\sum_zP_z{\ket{z}}\bra{z}\otimes \rho^z_E$ with normalized $\rho^z_E$. Eve's knowledge about the
possible outcomes of the $\mathbb Z$ measurements is given by the min-entropy $H_{\rm min}(Z|E)$, evaluated over $\rho_{ZE}$.
If Alice sometimes measures her system with a different POVM $\mathbb X$, the UP allows to bound
the min-entropy $H_{\rm min}(Z|E)$ and then the guessing probability by eq. \eqref{p_guess}. 
In fact, by using eq. \eqref{UP} and by considering
the system $B$ as a trivial space, the uncertainty relation becomes
$H_{\text{min}}(Z|E)\geq q-H_{\text{max}}(X)$, 
where the max-entropy must be evaluated on the state obtained by the $\mathbb X$ measurement, namely
$\rho_{X}\equiv\sum_x p_x\ket{x}\bra{x}$, 
with $p_x=\text{Tr}_{AE}[\N_x\rho_{AE}]$.
{In this case $H_{\text{max}}(X)=2\log_2\text{Tr}[\sqrt{\rho_X}]$
(see Appendix \ref{min-entropy} and \cite{koni09ieee})}, i.e.  the max-entropy is equal to $H_{1/2}(X)$, the
R\'enyi entropy \footnote{We recall that the R\'enyi entropy of order $\alpha$
is defined as $H_\alpha(X)=\frac{1}{1-\alpha}\log_2\sum^{d-1}_{x=0}p^\alpha_x$.} 
of order $1/2$ of the classical outcome $X$.

Our result can be summarized as follows: the conditional min-entropy of the $\mathbb Z$ outputs
can be bounded by using the R\'enyi entropy of order $1/2$ of the  $\mathbb X$ outputs, namely 
\beq
\label{main_result}
H_{\text{min}}(Z|E)\geq q-H_{1/2}(X)\,.
\eeq
{that reduces to \eqref{simple_bound} in case of conjugate observables in $d$ dimensions.}
We would like to point out that, thanks to the inequality $H_{1/2}(X)+H_{\infty}(Z)\geq q$ derived by Maassen
and Uffink \cite{maas88prl}, the bound $q-H_{1/2}(X)$ is always lower than the classical min-entropy $H_{\rm \infty}(Z)$
evaluated on the probabilities $P_z$.

 \begin{figure}[tbp] 
\begin{center}
\includegraphics[width=8.5cm]{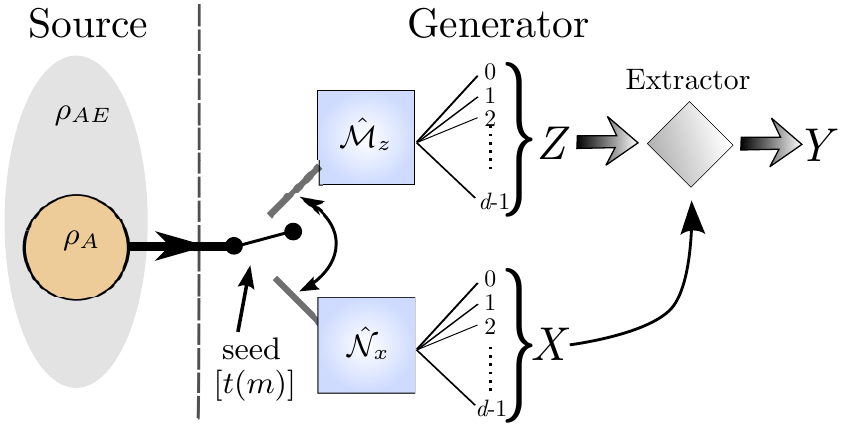}
\caption{(Color online) Scheme of the QRNG. The source of randomness is the state $\rho_A$ that can be correlated with a larger system $E$. 
An initial perfect random seed of length $t(m)$ is used to switch
between the $\{\M_z\}$ and $\{\N_x\}$ POVMs, from which the random variables $Z$ and $X$ are extracted.
The variable $Z$ is used to generate the random sequence,
while the variable $X$ is used to evaluate how many true random bits can be extracted by $Z$. $Y$ represents the final true random sequence.}
\label{fig1}
\end{center}
\end{figure}

\section{UP-certified QRNG} 
Let's now evaluate the bound 
 in two particular cases.
Let's consider the $\mathbb Z$ POVM as projective 
measurements in the computational basis, $\{\ket{0}$, $\ket{1},\cdots,\ket{d-1}\}$
and the $\mathbb X$ measurement chosen as its discrete-Fourier transform 
$\ket{x}=\frac{1}{\sqrt{d}}\sum^{d-1}_{z=0}e^{i \frac{x z}{2\pi d}}\ket{z}$ for which $q=\log_2d$. 
If the system $A$ is prepared in the 
state $\ket{\psi}_A=\frac{1}{\sqrt{d}}\sum_z\ket{z}$, then 
 $H_{1/2}(X)=0$ and {\eqref{main_result} bounds $H_{\text{min}}(Z|E)$ to the classical min-entropy $H_{\infty}(Z)=\log_2d$}.
The random variable $Z$ is then uniformly distributed and independent from any adversary. 
However, in practical implementations of a QRNG, it is impossible to prepare
the system $A$ in a perfect pure state $\ket{\psi}_A$. 
{When the state $\rho_A$ is not pure, the entropies  $H_{\infty}(Z)$ and  $H_{\rm min}(Z|E)$  can be different.
Our result is thus particularly effective with real sources (that cannot generate pure states) since 
it bounds the effective achievable randomness without requiring any assumption on them}. 
Even if Eve has complete control on the source $\rho_A$, the bound given in \eqref{main_result} evaluates the amount of  
true random bits that can extracted from $Z$. 
This randomness has complete quantum origin and no side information can be used to predict the generated random bits.

Another important example is represented by the system described in the introduction: Eve sends to Alice one photon of a 
two-photon maximally entangled state, and thus can perfectly predict the outputs of Alice's measurements.
In this case, Alice holds a completely mixed state $\rho_A=\frac12\openone_2$
and  the max-entropy is $H_{1/2}(X)=1$. Thanks to eq. \eqref{main_result} and \eqref{p_guess}, 
the bound on the min-entropy becomes trivial, $H_{\rm min}(Z|E)\geq 0$ and $p_{guess}(Z|E)\leq 1$:
our result correctly predicts that the guessing probability can reach unity and so no true random bits can be extracted in this case.

In order to exploit the result of eq. \eqref{main_result} it is necessary to estimate the max-entropy of the source $\rho_A={\rm Tr}_E[\rho_{AE}]$.
However, since the POVM $\{\M_z\}$ and $\{\N_x\}$ are incompatible, it is not possible to measure them at the same time.
We then need to switch randomly between $\M_z$ and $\N_x$ during the random bit generation (see Figure \ref{fig1}). 
The measurements are chosen by using a seed of true randomness that our method is able to expand. 
From this point of view, our method can be seen as a random number expansion protocol.

We now show that the number of random extracted bits is greater than the required seed. 
 Let $m$ the total number of measurements. 
 We decide that, over $m$, the number of measurements in the  POVM $\{\N_x\}$ will be $n_X=\lceil\sqrt{m}\rceil$, such that
 the probability of measuring in the $\mathbb X$ basis is  approximately $\frac{1}{\sqrt{m}}$. 
 To randomly choose $n_X$ among $m$ measurements we need a number of bits given by
 $t(m)=\lceil\log_2\frac{m!}{n_X!(m-n_X)!}\rceil$. This is the length of the random seed required for the randomness expansion.
 
The  probabilities of outcomes in the $\mathbb X$ basis are given by $p_x={\rm Tr}_{A}[\N_x\rho_{A}]$
and the asymptotic lower bound of the min-entropy is $H_{\rm min}(Z|E)\geq q-H_{1/2}(X)$.
 From the experimental point of view we need to estimate the max-Entropy $H_{1/2}(X)$ by using the $n_X$ outcomes.
If we denote by $n_x$ the number of outcomes such that $X=x$, we can estimate the max-entropy  
by using the Bayesian estimator
 defined in \cite{hols98jpa} {(with a uniform prior distribution)}:
 \beq
 \widetilde H_{1/2}(\{n_x\})=2\log_2[\frac{\Gamma(n_X+d)}{\Gamma(n_X+d+\frac12)}
 \sum^{d-1}_{x=0}\frac{\Gamma(n_x+\frac32)}{\Gamma(n_x+1)}]\,.
 \eeq
The Bayesian estimator has a lower variance with respect to the 
frequentist estimator $ \widetilde H^f_{1/2}=2\log_2[\sum^{d-1}_{x=0}\sqrt{\frac{n_x}{n_X}}]$. Moreover, for low max-entropies,
the frequentist estimator has a negative bias that overestimates the bound on the min-entropy.

Then, given $m$ measurements, the number of extracted random bits are the outputs of the $\mathbb Z$ measurement,
given by $m-n_X$: due to the bound \eqref{main_result}, at least $(m-n_X)(q-H_{1/2}(X))$ are true random bits.
If we subtract the number of bits $t(m)$ required for the seed, 
we can estimate the random bits generation rate per measurement as
\beq
\widetilde r(\{n_x\})=\frac{b_{\rm sec}}{m}\,,
\eeq
 where $b_{\rm sec}$ is the number of generated true random bits :
\beq
b_{\rm sec}=(m-n_X)[q-\widetilde H_{max}(\{n_x\})]-t(m)\,.
\eeq
It is worth noticing that, in the infinite size limit $m\rightarrow+\infty$, the seed length is given by
 $t(m)\sim\sqrt{m}\log_2\sqrt{m}$,  the estimator  $\widetilde H_{1/2}(\{n_x\})\sim H_{1/2}(X)$, 
 and the rate approaches the asymptotic limit
$\widetilde r\xrightarrow{\quad} r(Z)=q-H_{1/2}(X)$.
Since the number of extracted random bits are quadratically larger than the initial seed bits, the generator
can work in loop: an initial seed is expanded and part of the extracted randomness is fed as a new seed.

\section{Experimental realization}
We have experimentally tested our method with
 two different random number generators implemented by
photon pairs generated in the $\ket{HV}$ state by spontaneous parametric down conversion.
See Appendix \ref{source} for details about the source.
The first generator is a single qubit QRNG, operated by an heralded single photon source: 
one photon of the pair, measured in the $\ket{H}$ state, is used as trigger, while
the second represents the signal.
By measuring the signal photon in the $\mathbb Z=\{\ket{+},\ket{-}\}$ and $\mathbb X=\{\ket{H},\ket{V}\}$ bases, 
we generate the random variables $Z$ and $X$. Here we denote with 
$\ket{\pm}$ the diagonal polarization states $\frac{1}{\sqrt2}(\ket{H}\pm\ket{V})$.
The second generator is a 4-level system (ququart) QRNG, represented by the pair of photons. In this case the $\mathbb Z$ and $\mathbb X$ bases
are respectively given by $\{\ket{++},\ket{+-},\ket{-+},\ket{--}\}$ and $\{\ket{HV},\ket{VV},\ket{HH},\ket{VH}\}$.

\begin{figure}[t]
\begin{center}
\includegraphics[width=8.5cm]{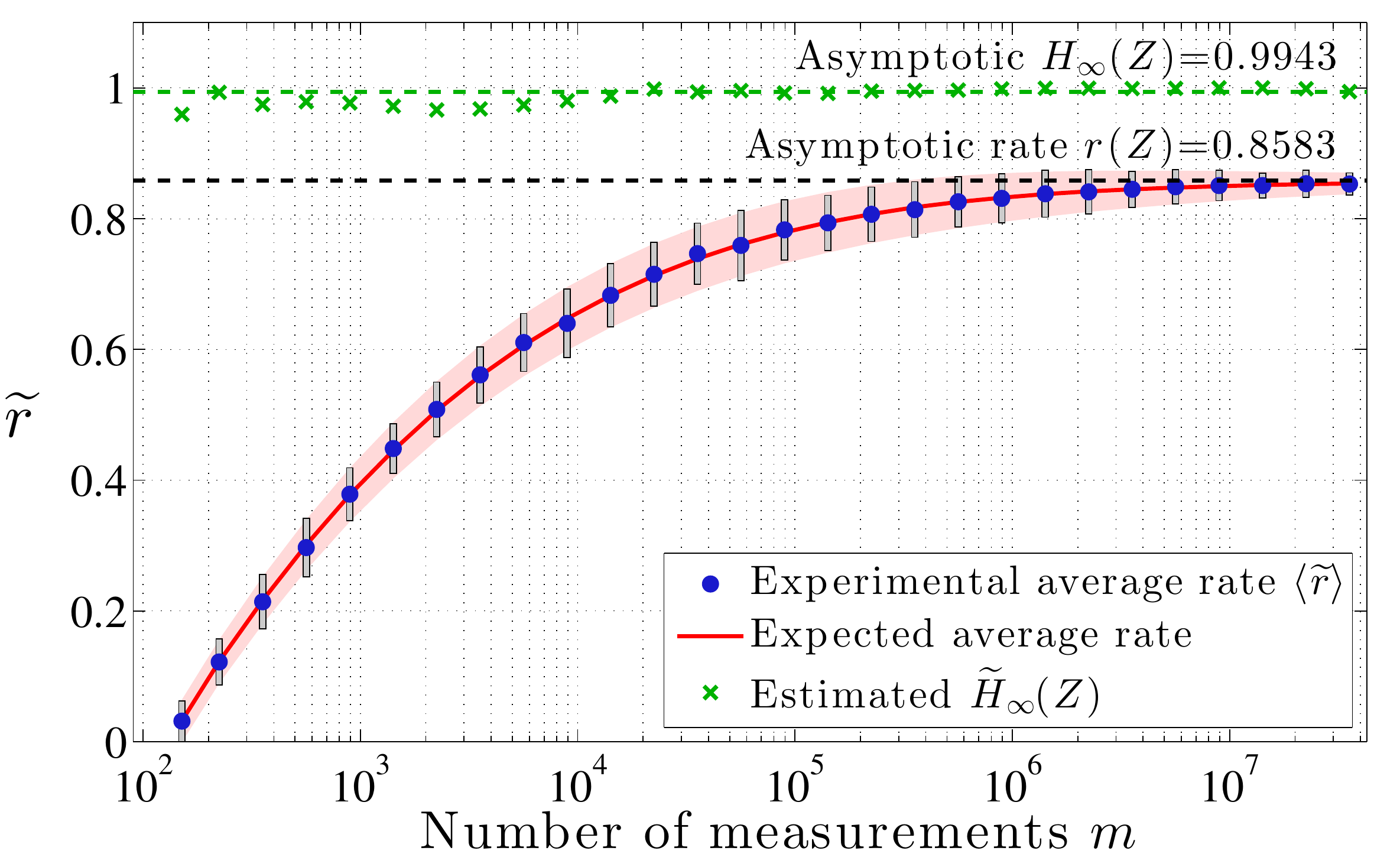} 
\caption{(Color online) Average experimental rate for the qubit QRNG. 
Blue circles represent the experimental average rate $\widetilde r$ of true random bits per measurement, while the 
continuous red line is the theoretical prediction with 
$\rho_X=\sum^1_{x=0}p_x\ket{x}\bra{x}$ where $p_0=0.9973$ and $p_{1}=0.0027$.
Shaded red area represents the theoretical standard deviation of the rate, while gray rectangles show the experimental standard
deviation of the rate.
Green crosses show the classical min-entropy estimated on the $Z$ random variable. The asymptotic limit $H_\infty(Z)$
is evaluated on the state $\rho_Z=\sum^1_{z=0}P_z\ket{z}\bra{z}$ with $P_0=0.5020$ and $P_{1}=0.4980$.}
\label{qubit}
\end{center}
\end{figure}

We first analyze the qubit QRNG. By choosing different values of $m$ we performed $n_X=\lceil\sqrt{m}\rceil$ measurements in the
$\mathbb X$ basis and $n_Z=m-n_X$ measurements in the $\mathbb Z$ basis, obtaining the sequences $X$ and $Z$.
The two sequences are used to estimate the classical max-entropy $\widetilde H_{1/2}(\{n_x\})$
and the rate $\widetilde r(\{n_x\})$. For each $m$, in figure \ref{qubit} we show the average 
rate $\widetilde r$ and its standard deviation experimentally evaluated  over 200 different $X$ sequences of $n_X$ bits
(see Appendix \ref{rate_analysis} for the rate achieved, for each $m$, by a single $X$ sequence of $n_X$ bits).
The experimental rates can be compared with the predicted average rate
$\langle\widetilde  r\rangle=\sum_{\{n_x\}}\Pi(\{n_x\})\widetilde r(\{n_x\})$, obtained by averaging $\widetilde r(\{n_x\})$
over the multinomial distribution
$\Pi(\{n_x\})=\frac{n_X!}{n_0!n_1!\cdots n_{d-1}!}p^{n_0}_0p^{n_1}_1\cdots p^{n_{d-1}}_{d-1}$.
We also show the classical min-entropy $\widetilde H_\infty(Z)$ evaluated on a sequence $Z$ with $n_Z$ bits.
The figure shows a very good agreement between the experimental result and the theoretical prediction. It is worth noticing
that at least $m>150$ measurements are necessary to obtain a positive rate $\widetilde r$, while with just $m\simeq 10^6$ the 
rate is very close to the asymptotic bound $r(Z)$.
The difference between $H_\infty(Z)$ and $\widetilde r$ corresponds to the possible knowledge that an adversary
holding the system $E$ may have. 
The limit $H_\infty(Z)$ is often and erroneously taken as the amount of true randomness used to calibrate the extractor:
in this way, even if the output string appears statistically good, possible side information held by Eve is not completely erased.
In our experimental analysis, since we are mainly interested
in demonstrating the physical principles,
we did not use active switches to change between the two bases (we first measured the $Z$ sequence and afterwards
the $X$ sequence). For practical applications, however, the QRNG should contain an active switch controlled by the seed $t(m)$.

\begin{figure}[tbp]
\begin{center}
\includegraphics[width=8.5cm]{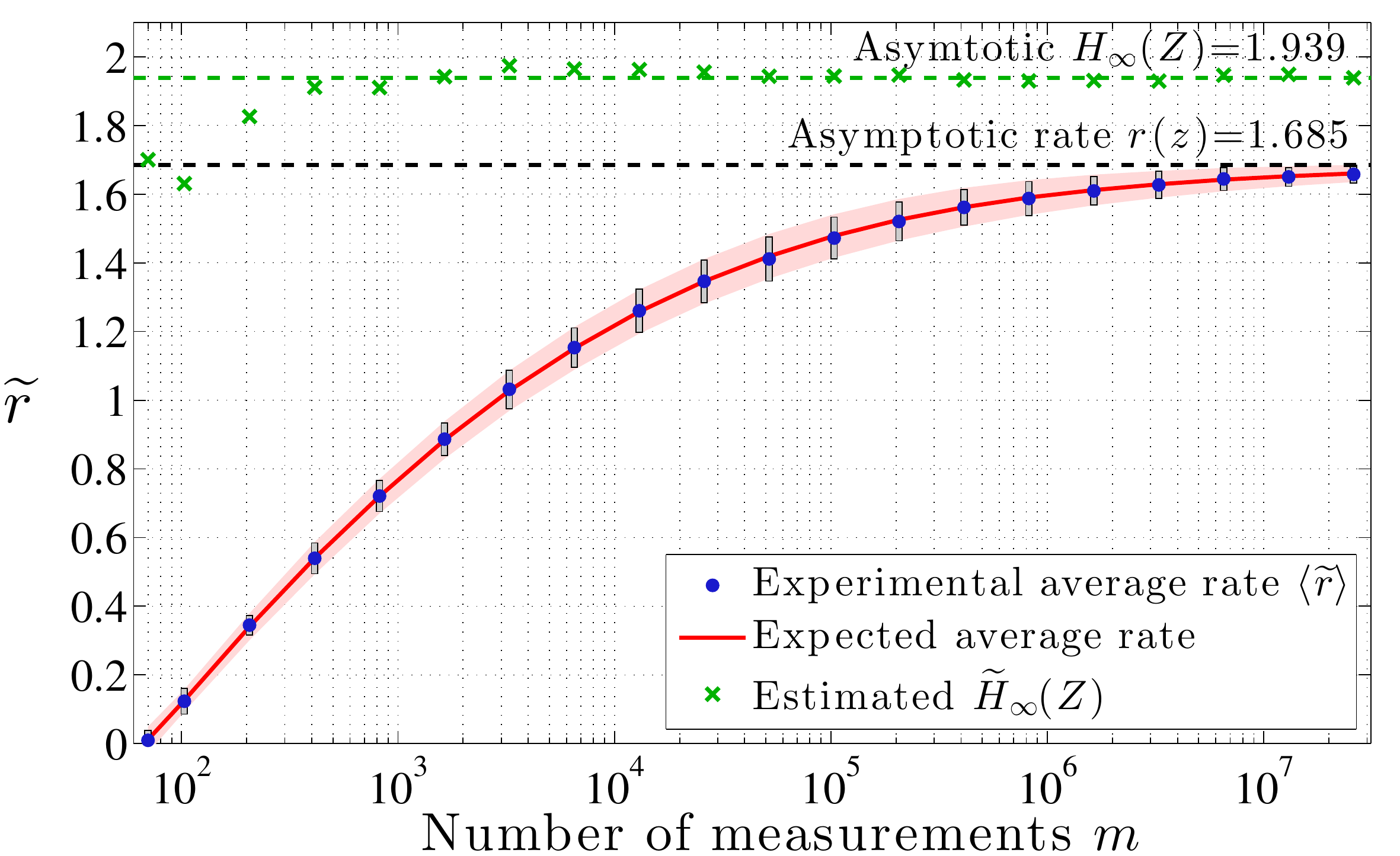}
\caption{(Color online) Average experimental rate for the ququart QRNG. See figure \ref{qubit} for notations.
In this case
$\rho_X=\sum^3_{x=0}p_x\ket{x}\bra{x}$ with $p_0=0.9937$, $p_{1}=0.00359$, $p_{2}=0.00266$ and $p_3=1-p_0-p_1-p_2$
and $\rho_Z=\sum^3_{z=0}P_z\ket{z}\bra{z}$ with $P_0=0.2527$, $P_{1}=0.2412$, $P_{2}=0.2608$ and $P_{3}=0.2453$.}
\label{ququart}
\end{center}
\end{figure}

In figure \ref{ququart} the results for the ququart QRNG are presented. Also in this case,
 for each $m$, the average 
rate $\widetilde r$ and its standard deviation are experimentally obtained by 200 different $X$ sequences of $n_X(m)$ bits.
Again, there is a very good agreement between the experimental results and the theoretical predictions and
a positive (average) rate is obtained for $m>70$. As before, for $m\simeq 10^6$ the rate is very close to the asymptotic bound $r(Z)$:
thanks to the larger Hilbert space, we can asymptotically obtain $1.685$ bits per measurement, that should be compared
with the value $0.8583$ achieved with the qubit QRNG. 
Our method is thus very robust 
with respect to the increasing of the dimension $d$ of the system.

For the complete proof of our protocol, we 
performed the extraction on a long random sequence $Z$ and the results are presented in Appendix \ref{tests}.

\subsection{Detailed comparison with Ref. \cite{fior07pra}}
Here we give a detailed comparison between our method and the result of Fiorentino {\it et al} \cite{fior07pra},
where the conditional min-entropy of a qubit state is evaluated by measuring its density matrix 
$\rho=\frac{1}{2} (\openone+\vec r\cdot\vec\sigma)$ ($\sigma_i$'s are the Pauli matrices and $\vec r$ is a three-dimensional vector
such that $|\vec r|\leq 1$). By extracting the random bits by measuring the qubit in the computational basis $\mathbb Z=\{\ket{0},\ket{1}\}$ 
such that $r_z=\bra{0}\rho\ket{0}-\bra{1}\rho\ket{1}$, 
the conditional  min-entropy was estimated to be $H_{\rm min}(Z|E)=1-\log_2(1+\sqrt{1-{r^2_x-r^2_y}})$~\cite{fior07pra}.

 Our method estimates the min-entropy of the $Z$ outcomes by measuring in the
 $\mathbb X=\{\ket{\pm}\}$ basis giving the asymptotic bound of $H_{\rm min}(Z|E) \geq 1-\log_2(1 + \sqrt{1-r^2_x})$. 
 Our result is a lower bound, since $q-H_{1/2}(X) =
1-\log_2[1+\sqrt{1-{r^2_x}}]$:
the bound is tight when $r_y=0$. If the state is pure, the result of \cite{fior07pra} allows to achieve
the upper limit $H_{\rm min}(Z|E)=H_{\infty}(Z)$.  
The advantage of our approach resides in the fact that it is not necessary to measure
the full density matrix but only measurements on two mutually-unbiased basis.
Indeed, in order to evaluate the density matrix,
it is necessary to measure the system also in the $\mathbb X$ and $\mathbb Y=\{\frac1{\sqrt2}(\ket0\pm i\ket1)\}$ basis
 beside the basis chosen to obtain the random sequence.
Also in the case of \cite{fior07pra}, a random seed is needed to switch between the tomography bases and the random sequence basis. 
As a final consideration, the result of Fiorentino {\it et al.} applies only to qubit systems, while our result can be applied to a general qudit systems,
as we have demonstrated by analyzing the ququart QRNG. 

We now give a detailed comparison for finite $m$: let's
consider the following parameters $r_z=0.9947\pm0.001$ and $r_x=0.004\pm0.002$ corresponding
to the experimental measured parameter of our qubit QRNG. Since the norm of the vector $\vec r$ cannot be greater that 1, it implies
that $|r_y|\leq\sqrt{1-r^2_z-r^2_x}\leq0.1027$ corresponding to a purity greater that  $\mathcal P_{\rm min}=0.9947$.
We recall that purity of the state $\rho$ is defined as $\mathcal P={\rm Tr}[\rho^2]=\frac{1+r_x^2+r_y^2+r_z^2}{2}$.
The measurement in the $Y$ basis will allow to determine the $r_y$ parameter. 

We performed the detailed comparison, in the finite $m$ case ($m$ is the total number of measurements), 
between our method and Ref. \cite{fior07pra}. To obtain a fair comparison we set $n^*_X=n^*_Y=\lceil\sqrt{m}/2\rceil$ as the number
of measurements in the $X$ and $Y$ basis respectively for the tomographic method of \cite{fior07pra}. 
Then the number of measurements in the $Z$ basis is given by 
$n^*_Z=m-2\lceil\sqrt{m}/2\rceil$. From such measurements the $r_x$ and $r_y$ parameters are estimated as (we used Bayesian estimators):
\beq
r_x=\frac{n_{0x}-n_{1x}}{n_{0x}+n_{1x}+2}
\,\qquad
r_y=\frac{n_{0y}-n_{1y}}{n_{0y}+n_{1y}+2}
\eeq
To randomly choose the $X$ and $Y$ measurements over the total number of measurements $m$ we need a number of bits
given by $t^*(m)=2\lceil\log_2\frac{m!}{(2n^*_X)!(m-2n^*_X)!}\rceil$.

In Fig. \ref{comparison} we show the comparison between the two rates in case of perfect pure state $\mathcal P=1$ and
in the case of $r_y=0$, corresponding to $\mathcal P=0.995$: 
the figure show that our results are slightly outperformed by the tomographic extractor only for high purity states $\mathcal P>0.995$ and in the large 
$m$ regime ($m>10^5$). A maximum of $15\%$ improvement with respect to the results shown in Fig. 2 is 
expected if the generated state is pure $\mathcal P=1$ and $N>10^8$. 
However, to obtain such limited advantage, a complication in the scheme, namely the measurement in the $Y$ basis, is required.

\begin{figure}[t]
\begin{center}
\includegraphics[width=8.5cm]{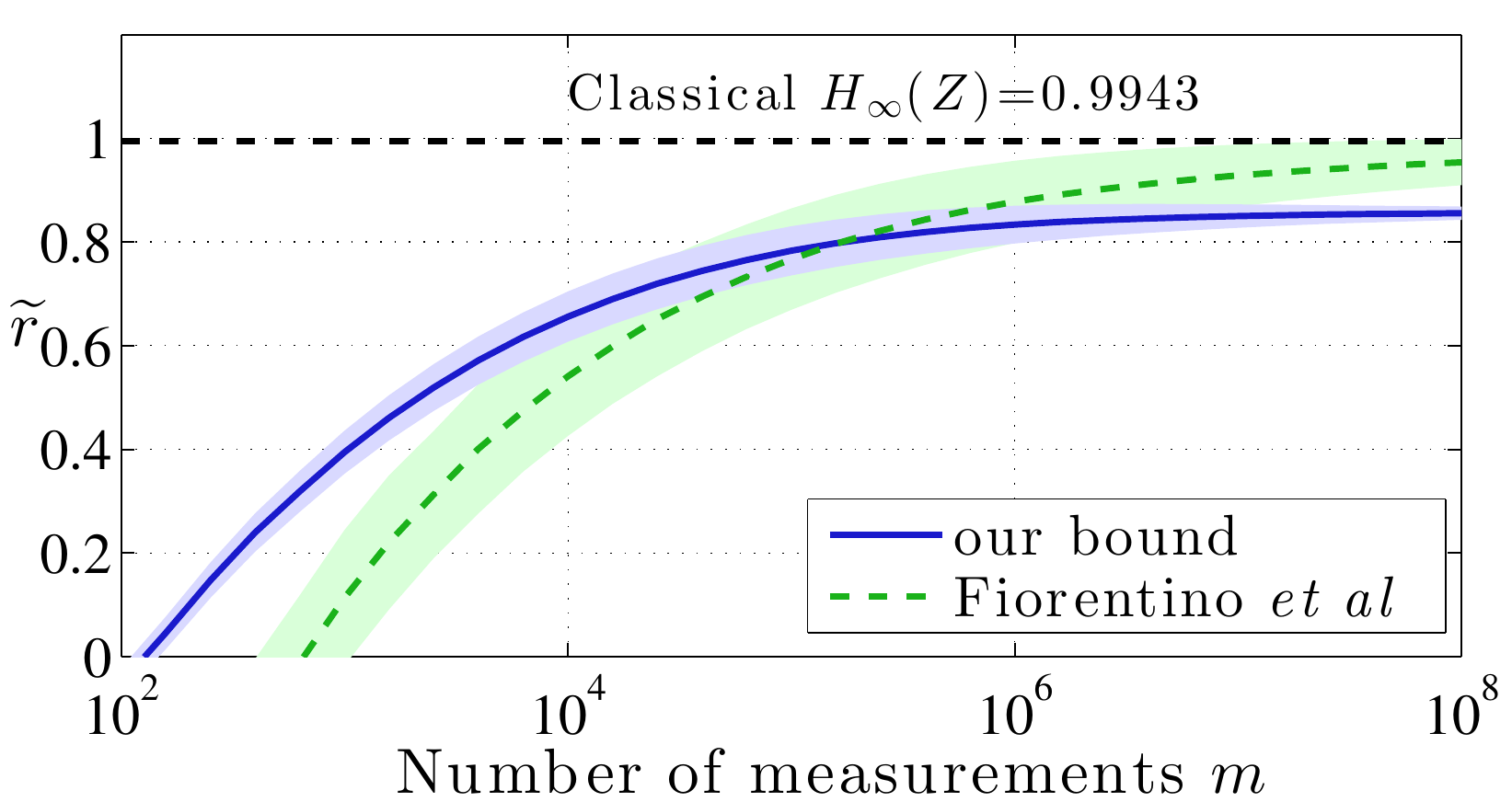}
\caption{(Color online) Comparison between the rate achievable by our bound (continuous blu line) 
and the rate achievable with the min-entropy estimation of Ref. \cite{fior07pra} (dotted green line)
in the case of perfect pure state with purity $\mathcal P=1$.}
\label{comparison}
\end{center}
\end{figure}

\begin{figure}[t]
\begin{center}
\includegraphics[width=8.5cm]{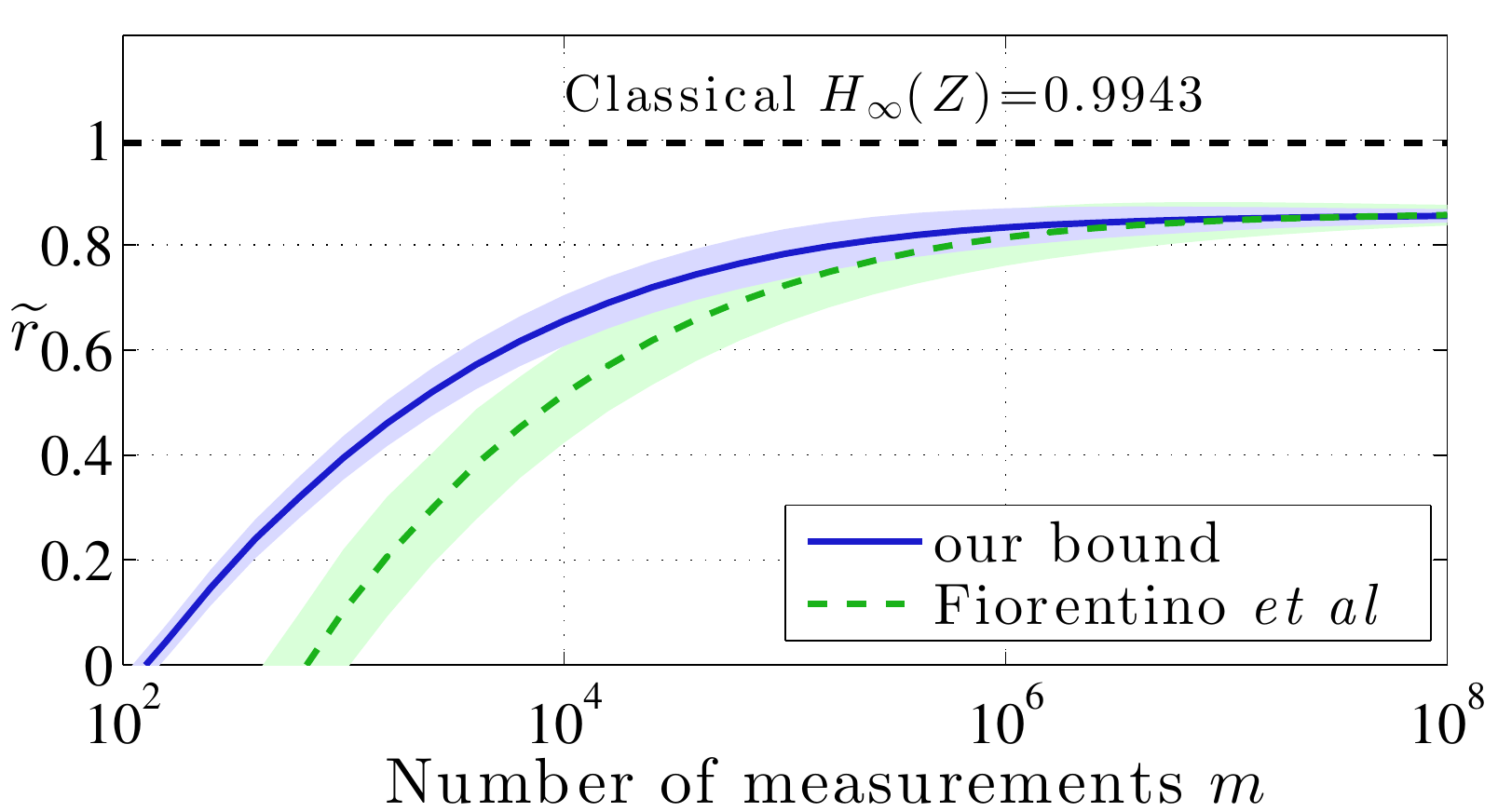}
\caption{(Color online) Comparison between the rate achievable by our bound (continuous blu line) 
and the rate achievable with the min-entropy estimation of Ref. \cite{fior07pra} (dotted green line)
in the case of slightly mixed state with purity $\mathcal P=0.995$.}
\label{comparison}
\end{center}
\end{figure}


\section{Conclusions}
We provided a bound, 
given by equation \eqref{main_result}, to directly compute the
conditional min-entropy $H_{\rm min}(Z|E)$ of the random variable $Z$, by using the classical random variable $X$.
The variables 
$Z$ and $X$ are obtained by measuring the system in two mutually unbiased bases. $H_{\rm min}(Z|E)$ represents
the amount of true 
randomness that can be extracted from $Z$.
No assumption is made on the source and/or
the dimension of Hilbert space. 
Our result is based on the fact the measurement device is trusted:
we assumed that the measurement system (waveplates and PBSs) works properly and the detector efficiency is not 
dependent on the input state or on an external control. 
{In order that detection system is only sensitive to a well known and characterized 
finite dimensional subspace of the total Hilbert space, photon number resolving detectors or the squashing model of QKD 
\cite{beau08prl,gitt14pra} can be implemented.}
It is important to stress that if the source does not generate a perfect pure state (and this always happens in experimental
realizations), the randomness extracted by standard methods, namely by measuring the system in a single basis,
is not a true randomness: an eavesdropper can have (partial or full) information about the generated random bits.
We have also tested our bound with a qubit and a ququart QRNG with good agreement
between theory and experiment.

Our method can be extended by taking into account possible imperfections in the measurement device, as illustrated in \cite{frau13qph}.
We believe that our method can be very useful for the extraction of true randomness and can be
applied in the framework of practical high-speed QRNG \cite{jofr11ope,abel14ope}, since it guarantees protection against quantum
side information without the need of complex Bell violation experiment.

\acknowledgments
We would like to thank Alberto Dall'Arche of the University of Padova for his support on the setup preparation.
Our work was supported by the Strategic-Research-Project QUINTET of the Department of
Information Engineering, University of Padova and the Strategic-Research-Project QUANTUMFUTURE (STPD08ZXSJ) of the University of
Padova.

\appendix

\section{Min and Max-entropy}
\label{min-entropy}
We here briefly review the definition of conditional min- and max- entropies introduced in \cite{koni09ieee}. 
The conditional min-entropy of a bipartite quantum state $\rho_{AE}$ is defined as:
 \beq
\label{min-entropy}
 H_{\rm min}(A|E)_{\rho_{AE}}=
 \max_{\sigma_B}{\sup}\Big\{\lambda\in \mathbb R\Big|\frac{\openone_A\otimes\sigma_E}{2^\lambda}\geq\rho_{AE}\Big\}\,,
 \eeq
where $\sigma_E$ is a normalized positive state.

The conditional max-entropy is the dual of the min-entropy. In fact, by using a purification
 $\rho_{ABC}$ of $\rho_{AB}$, the max-entropy is defined by
 \beq
\label{max-entropy}
 H_{\rm max}(A|B)_{\rho_{AB}}=- H_{\rm min}(A|C)_{\rho_{AC}}\,,
 \eeq
where $\rho_{AB}={\rm Tr}_C[\rho_{ABC}]$ and $\rho_{AC}={\rm Tr}_B[\rho_{ABC}]$.
We here recall that
the  purification of a state $\rho_{AB}$ is a pure state $\rho_{ABC}$ 
in the extended Hilbert space $A\otimes B\otimes C$,
such that ${\rm Tr}_C[\rho_{ABC}]=\rho_{AB}$.

For the QRNG we need to evaluate the max-entropy for the state $\rho_{X}\equiv\sum^{d-1}_{x=0} p_x\ket{x}\bra{x}$, where the space $B$
is a trivial space. By definition \eqref{max-entropy} we have:
\beq
H_{\rm max}(X)_{\rho_X}=-H_{\rm min}(A|C)_{\rho_{AC}}
\eeq
with $\rho_{AC}$ a purification of $\rho_{X}$. A possible purification is given by
\beq
\rho_{AC}=\ket{\Psi}_{AC}\bra\Psi\,,\qquad
\ket\Psi_{AC}=\sum^{d-1}_{x=0}\sqrt{p_x}\ket{x}_A\otimes\ket{v_x}_C
\eeq
with $\{\ket{v_x}\}$ on orthonormal basis on the space $C$ with dimension $d$. By \eqref{min-entropy} we have
\beq
\begin{aligned}
H_{\rm max}(X)_{\rho_X}&=-H_{\rm min}(A|C)_{\rho_{AC}}\\
&=- \max_{\sigma_B}{\sup}\Big\{\lambda\in \mathbb R\Big|\frac{\openone_A\otimes\sigma_C}{2^\lambda}\geq\ket{\Psi}\bra\Psi\Big\}\,,
\end{aligned}
\eeq

The state $\sigma_C$ that maximize min-entropy definition is $\sigma_C=\openone/d$. The maximum $\lambda$ such that
$\openone_A\otimes\openone_C\geq {d2^\lambda}\ket{\Psi}\bra\Psi$ is $\lambda=-\log_2[\sum_x(\sqrt{p_x})]^2$, such that
\beq
H_{\rm max}(X)_{\rho_X}=\log_2[\sum_x\sqrt{p_x}]^2=2\log_2\sum_x\sqrt{p_x}=H_{1/2}(X)
\eeq

\section{Photon source}
\label{source}
\begin{figure}[t]
\begin{center}
\includegraphics[width=8.5cm]{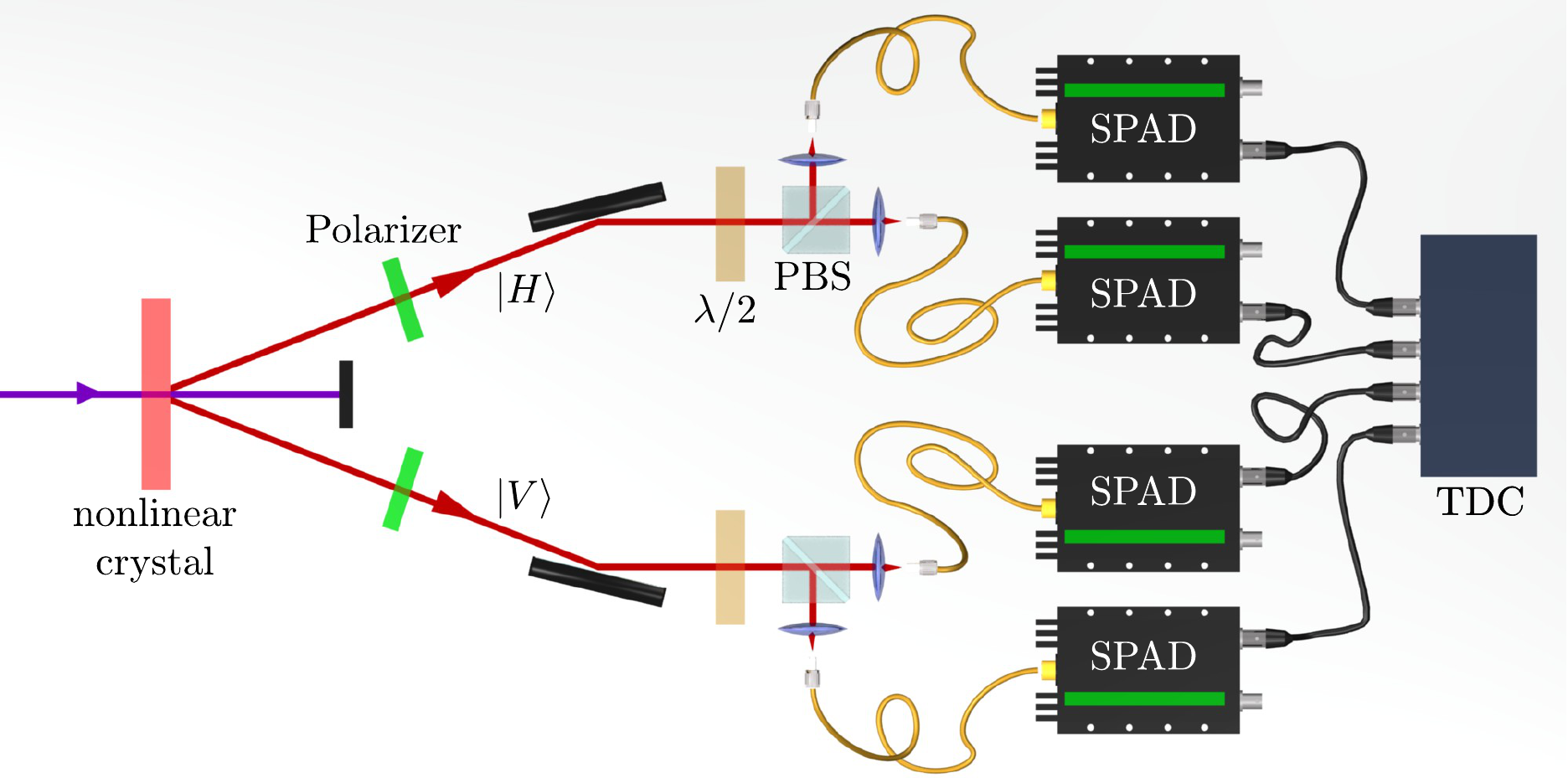}
\caption{(Color online) Scheme of the experimental setup generating the SPDC photons.
 TDC=Time to digital converter; SPAD=single-photon avalanche diode; PBS=polarizing beam splitter, $\lambda/2$=waveplates.}
\label{setup}
\end{center}
\end{figure}

Photons used in experimental demonstration of the method were generated by spontaneous parametric down conversion (SPDC),
as illustrated in figure \ref{setup}.
A femtosecond pulsed laser (76MHz repetition rate) at 405nm shines a nonlinear crystal where pairs of photons
are probabilistically emitted over two correlated directions. Two polarizers select the $\ket{HV}$ pair, with
$\ket{H}$ and $\ket{V}$ the horizontal and vertical polarized photon respectively. Half waveplates $\lambda/2$
allow to change between the $\mathbb Z$ and $\mathbb X$ basis. For the single qubit QRNG, the $\ket{H}$ photon 
is used as trigger: its detection heralds the presence of the $\ket{V}$ photon. Single photon detectors (SPAD) deliver signals 
to a time-to-digital converter (TDC).
Concerning the rate of raw bits extraction, the source has a coincidence rate of 12 kHz:
we would like to point out that we are not interested in the speed of the generator, 
but on the demonstration of the method here presented. 
However, it is worth noticing that sources producing photon pairs at the rate of few MHz are currently available \cite{stei12ope, stei13ope}.

\section{Analysis of the random bit generation rate}
\label{rate_analysis}
In this section we show the experimental rate obtained with a single control $X$ sequence, while in the main text
we showed the average value obtained with 200 sequences. We report the rate achieved with the qubit QRNG.
We here recall that, given $m$ measurement on the state $\rho_A$, we obtained two classical $X$ and $Z$ sequences 
with $n_X$ and $n_Z$ bits respectively, whose lengths are respectively given by $n_X=\lceil\sqrt{m}\rceil$ and $n_Z=m-n_X$.
The state of the system $A$ after the measurement is given by $\rho_Z=\sum^1_{z=0}P_z\ket{z}\bra{z}$ or
$\rho_X=\sum^1_{x=0}p_x\ket{x}\bra{x}$, depending on the used POVM.

Given $m$, we would like to evaluate the "single shot" rate $\widetilde r$ given by:
\beq\label{rate_SI}
\widetilde r(n_0,n_1,m)=(m-n_X)(1-\widetilde H_{1/2}(n_0,n_1))-t(m)\,,
\eeq
with $n_0$ and $n_1$ the number of $0$'s and $1$'s in the $X$ sequence.

\begin{figure}[t]
\begin{center}
\includegraphics[width=8.5cm]{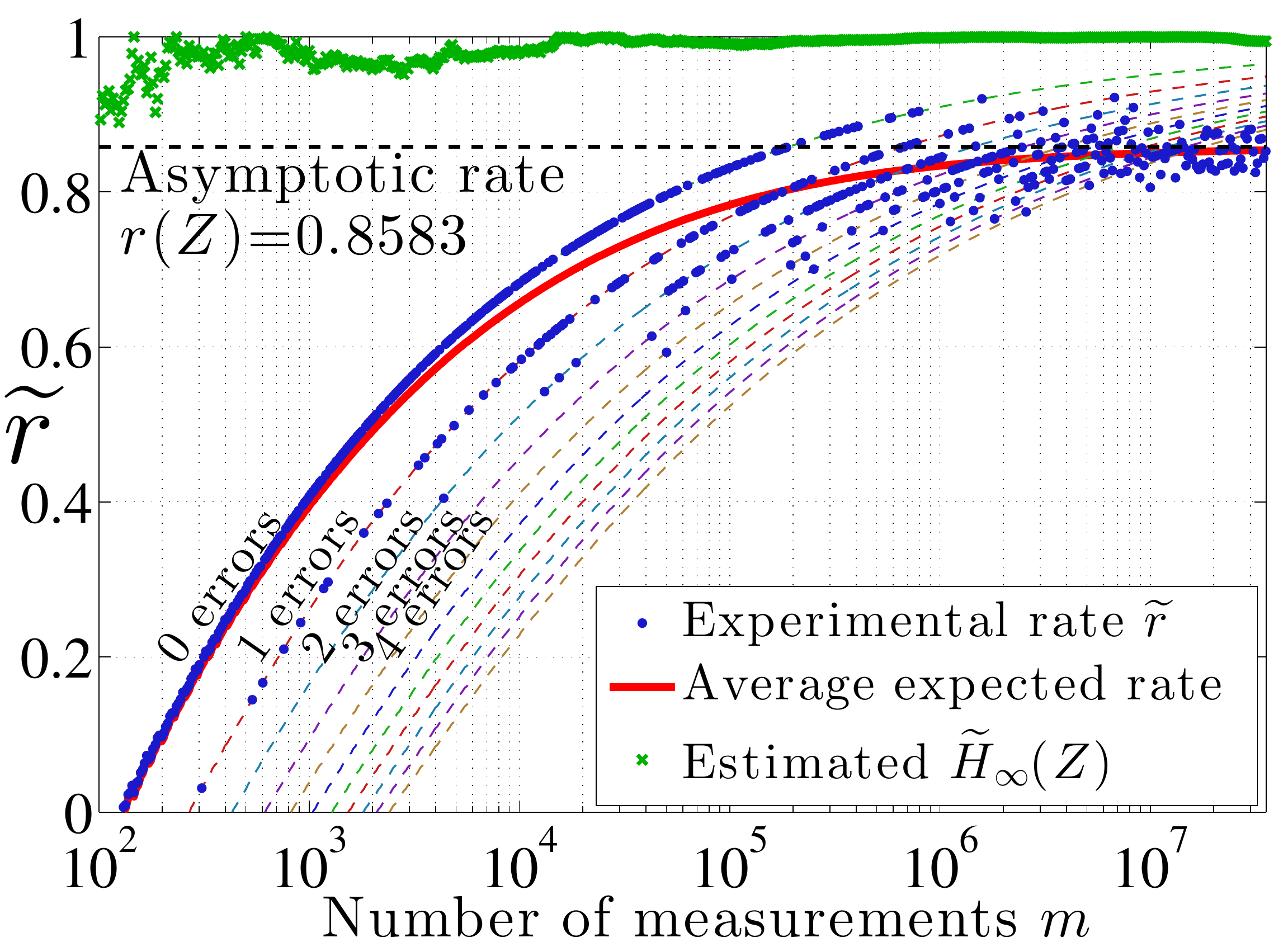}
\caption{(Color online) Experimental rate for the qubit RNG. 
Blu circles represents the experimental rate $\widetilde r$ of true random bits per measurement, while  
continuous red line represent the theoretical average prediction with 
$\rho_X=\sum^1_{x=0}p_x\ket{x}\bra{x}$ where $p_0=0.9973$ and $p_{1}=0.0027$.
Dashed lines represent the rate achieved with different number of "errors" in the $X$ sequence.
Green crosses show the classical min-entropy estimated on the $Z$ random variable obtained from the 
state $\rho_Z=\sum^1_{z=0}P_z\ket{z}\bra{z}$ with $P_0=0.5020$ and $P_{1}=0.4980$.}
\label{fig1supp}
\end{center}
\end{figure}
\begin{figure}[t]
\begin{center}
\includegraphics[width=8cm]{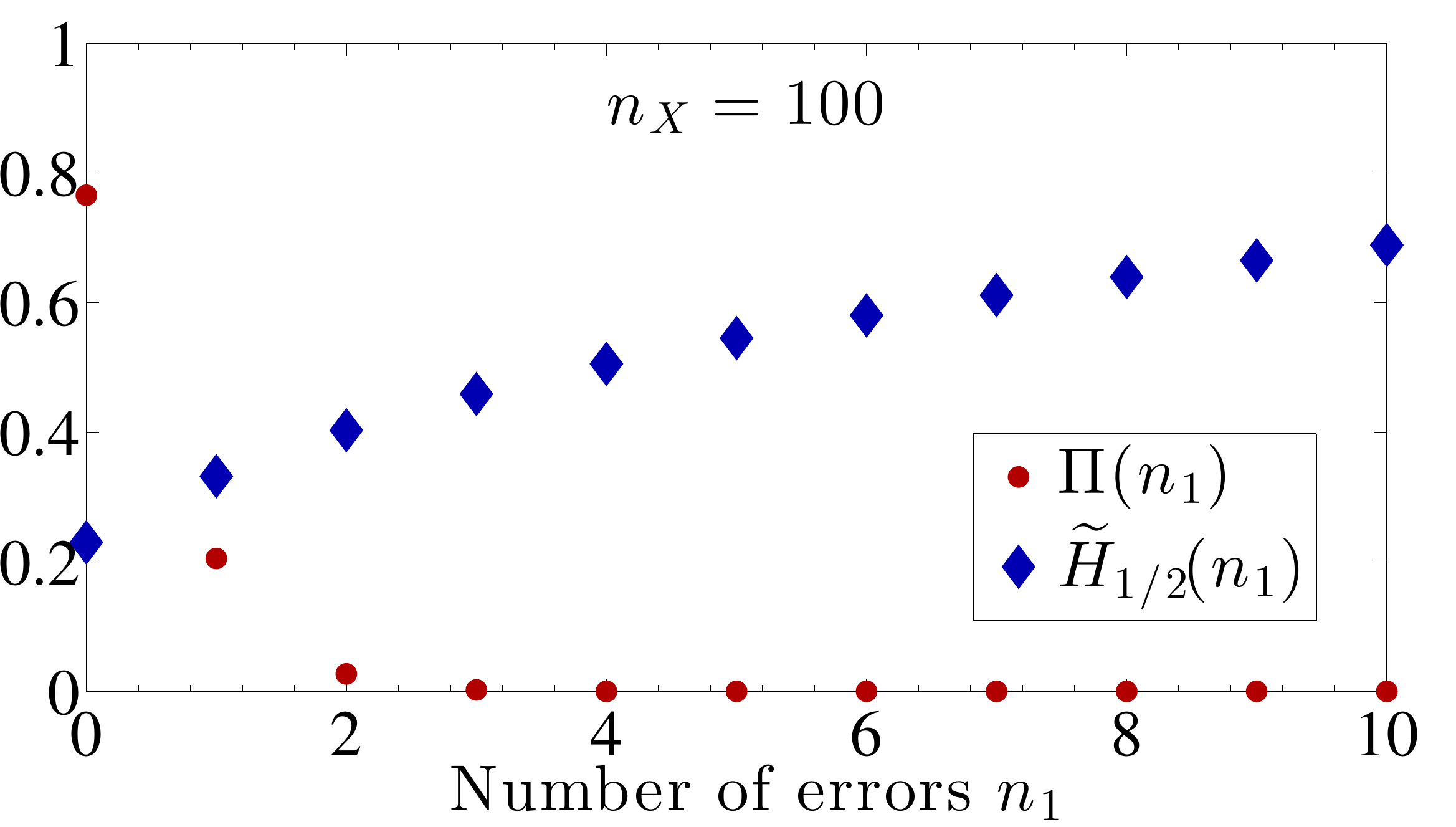}
\includegraphics[width=8cm]{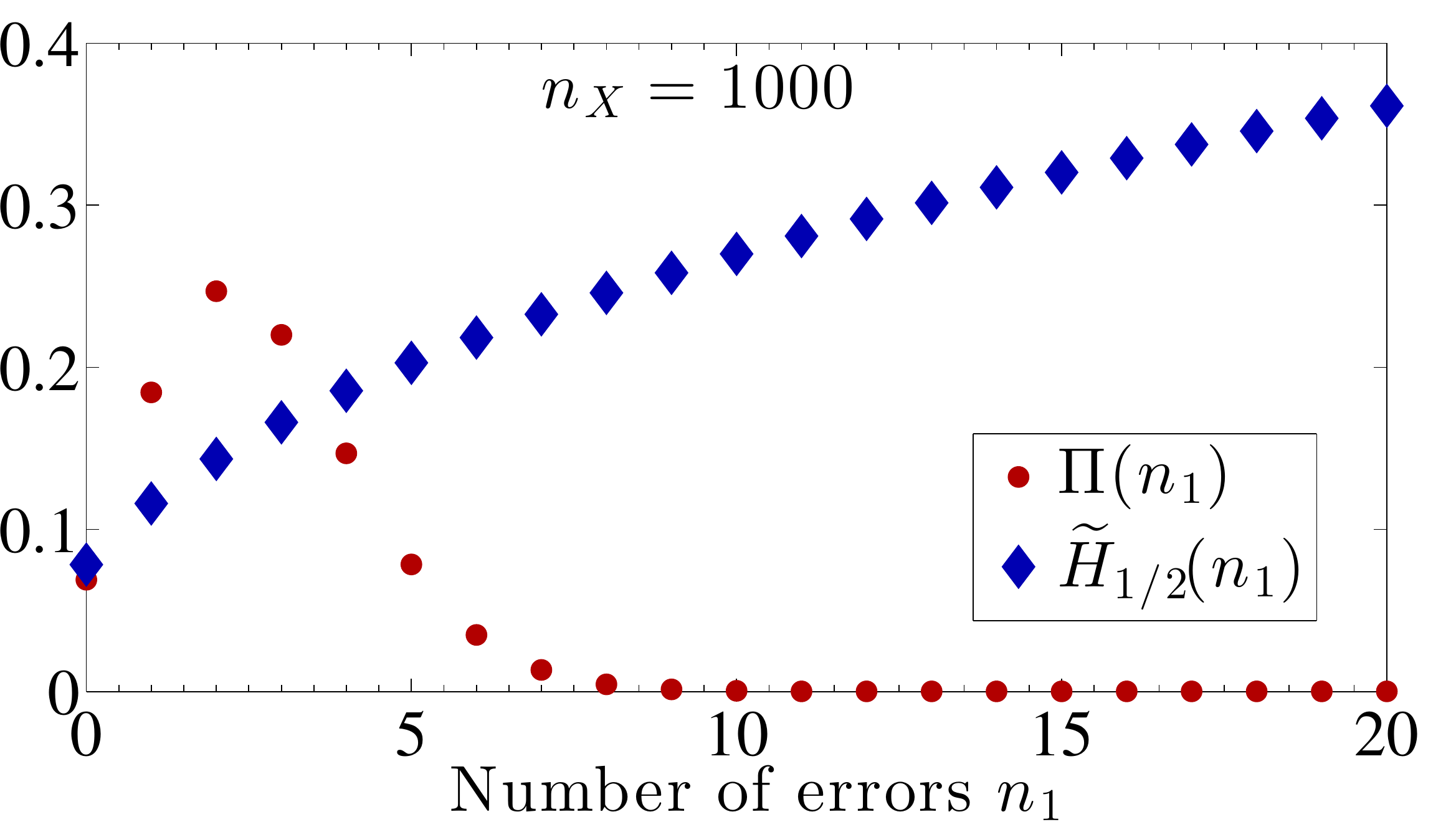}
\caption{(Color online) Extimated max-entropy $\widetilde H_{1/2}(X)$ and error probability $\Pi(n_1)$.
Due to the low value of $p_1=0.0027$, the $\Pi(n_1)$ is peaked around the low values of $n_1$.}
\label{fig2supp}
\end{center}
\end{figure}

 For the single qubit QRNG, since $n_0+n_1=n_X$, the single shot rate is function of only $m$ and $n_1$:
\beq
\begin{aligned}
\widetilde r(n_1,m)=&(m-n_X)\left\{1-2\log_2[\frac{\Gamma(n_X+2)}{\Gamma(n_X+\frac52)}]\right.
\\
&\left.-2\log_2[ \frac{\Gamma(n_X-n_1+\frac32)}{\Gamma(n_X-n_1+1)}+\frac{\Gamma(n_1+\frac32)}{\Gamma(n_1+1)}]\right\}
\\
&-\lceil \log_2\binom{m}{n_X}\rceil\,.
\end{aligned}
\eeq
For different values of $m$ we show in figure \ref{fig1supp} the achieved rate: 
each point represents the rate $\widetilde r$ evaluated over a single $X$ sequence of $n_X$ bits
obtained by the measurement in the $\mathbb X$ POVM. Each sequence is taken from a sample with the following property: 
\beq
\rho_X=\sum^1_{x=0}p_x\ket{x}\bra{x}\quad {\rm with}\quad  p_0=0.9973\,, p_{1}=0.0027.
\eeq
For perfect state preparation we would like to have $p_0=1$ and $p_1=0$: by this reason, the number of $1$ in the
$X$ sequence are defined as the "number of errors" in the sequence. The "errors" can 
be caused by the presence of the eavesdropper, or by imperfections in the preparation devices.
Since $p_1$ is very low, in Figure \ref{fig1supp}
it is possible to
see that, for $m<10^3$, few sequences have 1 errors and the most have 0 errors.
By increasing $m$, the number of errors increases to follow the prediction $n_1\sim p_1n_X$. 
For low $m$, the possible rates are "quantized", since the rate \label{rate_SI} is evaluated on  integer values $n_0$ and $n_1$.
In figure \ref{fig2supp} we show estimated max-entropy $\widetilde H_{1/2}(X)$ in function of the number of errors for the $n_X=100$
and $n_X=1000$ case. We also report the probability of obtaining $n_1$ errors, given by $\Pi(n_1)=\binom{n_X}{n_1}p_0^{n_0}p_1^{n_1}$.
The figure shows that $\widetilde H_{1/2}(X)$ has discrete values corresponding to different values of $n_1$.

\section{Tests on the extracted random numbers}
\label{tests}

\begin{table*}[!t]
\begin{center}
\includegraphics[width=8.9cm]{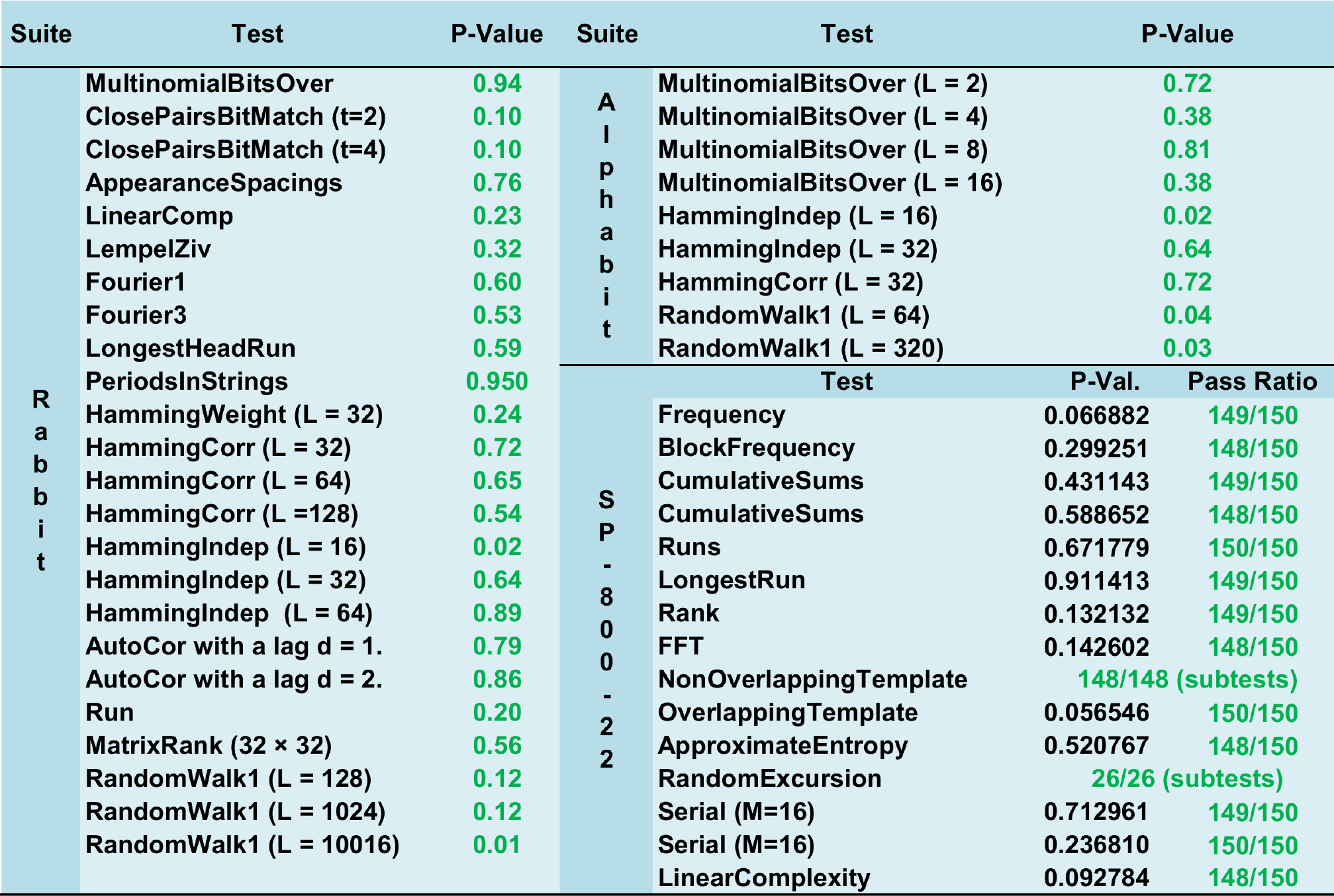}
\includegraphics[width=8.9cm]{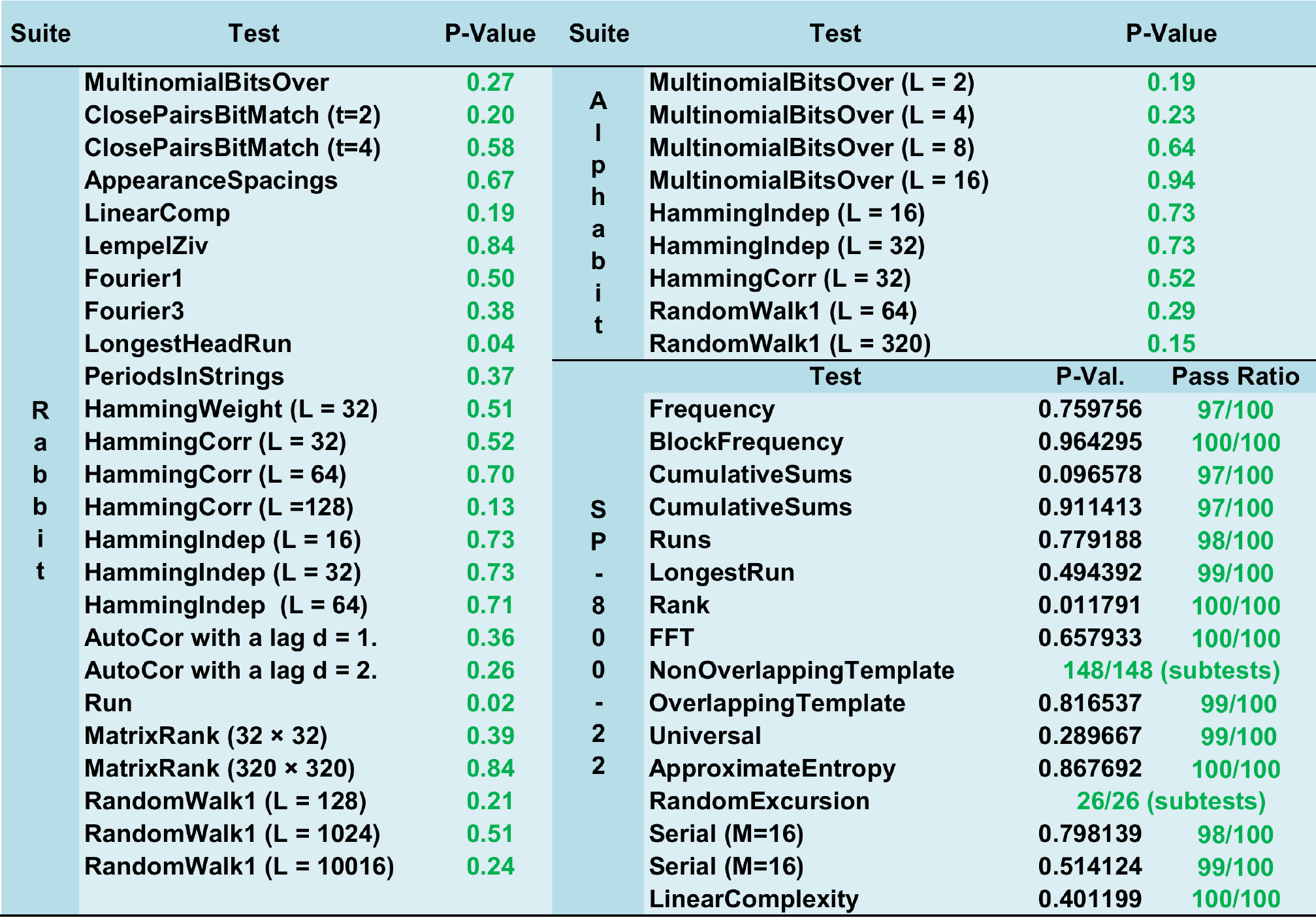}
\end{center}
\caption{(left) Summary of the results of selected tests of batteries  particularly effective in detecting defects in TRNG. The \textit{Alphabit} and \textit{Rabbit} batteries belong to the \textsc{TESTU01}: critical results are if $\mathcal{P}\text{-val} \leq 10^{-3}$ or $\mathcal{P}\text{-val} \geq 0.990$. For tests which give more than a p-values, the smallest is reported . For  NIST SP-800-22 suite, the file was partitioned in sub-strings $200 \,000$ bits long for a total of 150 strings: this length was chosen in order to obtain a sample sizes enough large such that it is  likely to fail the tests in case of poor randomness  with a significance level of $\alpha=0.01$; a test is failed if more than 6 strings fail it. In addition, a test is passed if the a chi-square test on the distribution of  p-values, gives it self a p-value $\mathcal{P}\text{-val} \geq 10^{-5}$.
(right) Summary of the results of selected tests of batteries  particularly effective in detecting defects in TRNG. The \textit{Alphabit} and \textit{Rabbit} batteries belong to the \textsc{TESTU01}: critical results are if $\mathcal{P}\text{-val} \leq 10^{-3}$ or $\mathcal{P}\text{-val} \geq 0.990$. For tests which give more than a p-values, the smallest is reported . For  NIST SP-800-22 suite, the file was partitioned in sub-strings $400 \,000$ bits long for a total of 100 strings: this length was chosen in order to obtain a sample sizes enough large such that it is  likely to fail the tests in case of poor randomness  with a significance level of $\alpha=0.01$; a test is failed if more than 4 strings fail it. In addition, a test is passed if the a chi-square test on the distribution of  p-values, gives it self a p-value $\mathcal{P}\text{-val} \geq 10^{-5}$.
 }
 \label{grid1}
\end{table*}

As a  quantitative example for the complete proof of our method, we 
performed the extraction on a long random sequence $Z$.
For the qubit case we use a random sequence $Z$ of length $n_Z=35.6\cdot 10^6$ and a control sequence $X$ of
length $n_X=5967$, requiring a seed length $t(m)=83443$. 
The estimated  lower bound for the min-entropy is $1-H_{1/2}(X) \simeq 0.8437$ giving an output random sequence $Y$ of
$b_{\rm sec}\simeq29.951\cdot 10^6$ bits.
For the qudit case, we have $ n_Z = 25.770\cdot 10^6$ and $n_X = 5100$ with a seed length $t(m)=70163$.
The estimated  lower bound for the min-entropy is $1.690$, giving $b_{\rm sec}\simeq43.886\cdot 10^6$ true random bits.
In both case, the initial $Z$ strings are fed to an extractor by two-universal hashing \cite{trev01acm,frau13qph} to obtain the $Y$ strings.
As we now will shown, the obtained bits pass successfully the most stringent tests \cite{rukh10web} for the  assessment of 
i.i.d. hypothesis for random bits.

At present time, the \textsc{TEST-U01}  \cite{TestU01} is the most stringent and comprehensive suite of tests; among all, we chose a pair sub-batteries, \textit{Rabbit} and \textit{Alphabit} respectively, specifically designed to tests RNGs.  The \textsc{SP-800-22} \cite{rukh10web} is developed by the \textit{NIST} and it is the most applied battery for RNG evaluation. 

The output of a test on a bit string is another random variable with a given distribution of probability, the so-called \textit{test statistic}. Hence, the $\mathcal{P} \text{-value}$, namely the probability of getting an equal or worse test statistic, holding true the i.i.d. hypothesis, are computed. If the $\mathcal{P} \text{-values}$ are smaller than some a priori defined critical values the tests are considered failed: these limits are usually chosen as $\mathcal{P} \text{-value} <0.01$ and $\mathcal{P} \text{-value} <0.001$, corresponding to a confidence level of 99\% and 99.9\% respectively. Otherwise, whenever one obtains $\mathcal{P} \text{-values}$ equal or greater than these limits,  the i.i.d. hypothesis for the tested string is assessed.

In Table \ref{grid1}  we report the results applied on the secure bits extracted by measuring a qubit and a ququart 
respectively. All the tests are passed.


%
%
\end{document}